# Site occupation of indium and jump frequencies of cadmium in FeGa$_3$

Randal Newhouse[1], Gary S. Collins[1], Matthew O. Zacate[2]

[1]Department of Physics and Astronomy, Washington State University, Pullman, WA, 99164, USA
[2]Department of Physics, Geology, and Engineering Technology, Northern Kentucky University, Highland Heights, KY 41099, USA

**Abstract**  Perturbed angular correlation (PAC) measurements using the In-111 probe were carried out on FeGa$_3$ as part of a broader investigation of indium site occupation and cadmium diffusion in intermetallic compounds. One PAC signal was observed with hyperfine parameters $\omega_1$ = 513.8(1) Mrad/s and $\eta$ = 0.939(2) at room temperature.  By comparison with quadrupole frequencies observed in PAC measurements on isostructural RuIn$_3$, it was determined that indium occupies only the 8j site in the FeGa$_3$ structure, denoted Ga(2) below because two out of the three Ga sites have this point symmetry.  PAC spectra at elevated temperature exhibited damping characteristic of electric field gradients (EFGs) that fluctuate as Cd probes jump among Ga(2) sites within the lifetime of the excited PAC level.  A stochastic model for the EFG fluctuations based on four conceivable, single-step jump-pathways connecting one Ga(2) site to neighboring Ga(2) sites was developed and used to fit PAC spectra.  The four pathways lead to two observable EFG reorientation rates, and these reorientation rates were found to be strongly dependent on EFG orientation.  Calculations using density functional theory were used to reduce the number of unknowns in the model with respect to EFG orientation.  This made it possible to determine with reasonable precision the total jump rate of Cd among Ga(2) sites that correspond to a change in mirror plane orientation of site-symmetry.  This total jump rate was found to be thermally activated with an activation enthalpy of 1.8±0.1 eV.

## 1    Introduction

Most intermetallic compounds are good conductors.  A few however, such as FeGa$_3$, RuGa$_3$ and RuIn$_3$ are actually small bandgap semiconductors [1,2].  The semiconducting nature of these compounds is thought to arise from greater than usual covalent bonding [3].  It is therefore of interest to investigate materials properties affected by cohesion in solids such as solute site occupation and diffusion jump rates.  The present work examined site occupation of indium and jump frequencies of cadmium in FeGa$_3$ using perturbed angular correlation spectroscopy (PAC).

The crystal structure of FeGa$_3$ (see Fig. 1) has been described in two different space groups [4, 5].  In the present work, the description from Ref. 5 is used.  Pairs of Fe atoms are found throughout a network of Ga atoms.  There are two symmetrically inequivalent Ga sites, labeled in the present work as Ga(1) and Ga(2) because there are twice as many Ga(2) as Ga(1) sites.  The Ga(2) sites form planes of atoms perpendicular to the c-axis of the tetragonal unit cell.  The Ga(1) sites and the Fe sites are contained within mixed planes midway between the Ga(2) planes.

One reasonably expects In to substitute for Ga because of their chemical similarity. Beyond that, however, it is not possible to predict with certainty site occupation of In among the two types of Ga sites. PAC using the [111]In probe can readily distinguish the distribution of In between the sites through the electric quadrupole interaction. Ga(1) and Ga(2) have different site symmetries which will lead to quadrupole interactions arising from electric field gradients (EFGs) with unique combinations of main principle component $V_{zz}$ and asymmetry parameter $\eta$. It is of interest to examine the In site occupation because a temperature dependence allows determination of the difference in site occupation energies, providing fundamental information about cohesion in the material as was the case, as examples, for $Pd_3Ga_7$ [6] and $Al_3Ti$ [7]. Even when PAC probe atoms are distributed across two sites belonging to different elements, information about the difference in site occupation energy can be obtained, though the measured energy also contains one or more intrinsic defect energies [8].

PAC is sensitive to the atomic scale motion of probes if a probe jumps among lattice sites with differing EFGs on a timescale that is comparable to the lifetime of the probe's PAC level (120 ns for [111]Cd, which is the daughter isotope of [111]In). Ga(1) has $2/m.$ site symmetry and Ga(2) has $.m$ site symmetry, which means that PAC probes will experience EFGs with differing orientations among sites of the same sublattice. Therefore, in addition to being able to detect shifts of probes back and forth among Ga(1) and Ga(2) sites, *i.e.* to inter-sublattice jumps, PAC will be sensitive to intra-sublattice jumps in this system. Changes in EFG strength and asymmetry and/or changes in EFG orientation accompanying probe jumps lead to a relaxation, or damping, of the PAC signal. In the slow fluctuation regime, the degree of damping is proportional to the total probe jump rate and has been used to characterize jump rates of the [111]Cd daughter isotope in a wide variety of intermetallic systems [6,9,10,11,12,13,14,15,16,17]

The rest of this paper is organized as follows. First, details about sample preparation and summaries of spectrometer setup and fitting methods are presented. Then, results of PAC measurements with fits of spectra using an empirical formula for nuclear relaxation are given. After that is an account of analysis of the nuclear relaxation using a stochastic model of fluctuating EFGs. This is followed by additional discussion of the findings.

## 2   Experimental Methods

### 2.1   Sample preparation

Samples of $FeGa_3$ were made by arc melting 99.999% purity Ga ingot and 99.9% purity Fe foil with carrier-free [111]In under Ar atmosphere. [111]In concentrations were very dilute at about $10^{-7}$ at.%. Samples were either annealed at 400 °C for one hour under about $3\times10^{-8}$ mbar vacuum, or the first PAC spectrum was collected under those conditions, effectively annealing the sample during the early stages of the first measurement. Signals from PAC spectra collected at room temperature exhibited only small frequency distributions, which indicated that the preparation method resulted in well crystallized and ordered crystals having low concentrations of point defects.

### 2.2   Perturbed Angular Correlation Spectroscopy

Measurements were made using a "slow-fast" PAC spectrometer with four fixed $BaF_2$ scintillation detectors separated by 90° in a plane. Measurements at elevated temperature were made in vacuum better than $10^{-6}$ mbar in a furnace designed to minimize sample-detector distance [18]. Spectra collected from samples at low temperature were fitted to a sum of static perturbation functions given by

$$G_2^{\text{static}}(\omega_Q, \eta, t) = s_0(\eta) + \sum_{n=1}^{3} s_n(\eta) \cos[\omega_n(\omega_Q, \eta) t] \tag{1}$$

where frequencies $\omega_n$ are functions of the asymmetry parameter $\eta$ and the quadrupole interaction frequency $\omega_Q \equiv |eQV_{zz}|/40\hbar$ and where the amplitudes $s_n$ are functions of $\eta$ [19]. At elevated temperatures, spectra exhibited relaxation characteristic of stochastically fluctuating EFGs. Such spectra were fitted both to an empirical formula and to an "exact" expression calculated numerically from a stochastic model for the EFG fluctuations [20,21].

The empirical formula used to fit spectra with damping due to dynamic effects is given by

$$G_2^{\text{dynamic}}(\lambda, \omega_Q, \eta, t) = \exp(-\lambda t) G_2^{\text{static}}(\omega_Q, \eta, t), \tag{2}$$

which is appropriate in the slow fluctuation regime where the relaxation parameter $\lambda$ is less than about $\omega_Q$ [22,23]. In this regime, $\lambda$ is inversely proportional to the average duration of an EFG experienced by a probe; when the EFG fluctuations are caused by probes jumping among lattice sites, then $\lambda$ is inversely proportional to the mean residence time of a probe in a lattice site.

In the present work, exact perturbation functions in the presence of stochastically fluctuating EFGs are calculated using

$$G_2(t) = \sum_q G_q \exp[(-\lambda_q + i\omega_q) t] \tag{3}$$

where $(-\lambda_q + i\omega_q)$ is the $q$th eigenvalue of the Blume matrix and $G_q$ are time-independent factors that depend on the eignevectors of the Blume matrix [20]. Terms in the Blume matrix depend on the EFG tensors experienced by probes and rates of transition among the EFGs [24]. Fits of spectra to the stochastic model were carried out using PolyPacFit [25] with stochastic models incorporated using the Stochastic Hyperfine Interactions Modeling Library [24,26].

## 3 Results

Previous studies of Cd movement in intermetallic compounds have shown there is value in studying relaxation rates for samples having compositions at opposing phase boundaries [9, 10, 27, 28, 29]; therefore, PAC measurements were made on samples prepared to be slightly rich and slightly deficient in Ga.

### 3.1 Ga-rich samples

Spectra collected from $FeGa_3$ samples prepared to be slightly Ga-rich showed only an unperturbed offset with no signal that could be attributed to indium in the $FeGa_3$ phase. According to the binary phase diagram, the only known phase richer in Ga than $FeGa_3$ is Ga metal [30]. It is therefore natural to attribute the unperturbed signal to indium in liquid Ga.

An X-ray diffraction measurement of one of the Ga-rich samples showed only the $FeGa_3$ phase, so any Ga metal present must have had a small volume fraction. It would therefore appear that In strongly segregates from FeGa$_3$ to the liquid Ga phase, which is consistent with other recent PAC studies that show In has a high affinity for liquid Ga [31, 32, 33].

3.2  Ga-poor samples

A typical PAC spectrum and its Fourier transform for a FeGa$_3$ sample prepared to be Ga-poor are shown in Fig 2. The three peaks in the Fourier transform correspond to a non-axially symmetric quadrupole interaction. While the measurement of the spectrum in Fig. 2 was collected at 400 °C, the room temperature quadrupole interaction parameters are $\omega_1$ = 513.8±0.1 Mrad/s and $\eta$ = 0.939± 0.002. This result indicates that $^{111}$In probe atoms occupy only one of the two available Ga sites. In order to identify which site, PAC spectra were collected for $^{111}$In in RuIn$_3$, which is isostructural with FeGa$_3$.

Since In is a host element in RuIn$_3$, the $^{111}$In must occupy both In sites in the stoichiometric 2:1 ratio. Accordingly, two signals were observed with a 2:1 ratio of amplitudes as can be seen by the spectrum measured at 500 °C and its Fourier transform shown in Fig. 2. When measured at room temperature, the two signals have hyperfine parameters $\omega_1$ = 415.8± 0.4 Mrad/s with $\eta$ = 0.015±0.015 and $\omega_1$ = 533.9±0.8 Mrad/s with $\eta$ = 0.9316±0.0006. The 534-Mrad/s signal has twice the amplitude as the other; therefore, the 534-Mrad/s signal is attributed to the In(2) site while the 416 Mrad/s signal is attributed to the In(1) site.

Since the PAC signal observed in FeGa$_3$ is very similar to the signal for the In(2) site in RuIn$_3$, both in fundamental frequency and asymmetry parameter, the 514-Mrad/s signal is attributed to probe atoms occupying the Ga(2) site in FeGa$_3$. These measurements therefore show that there is a very strong site preference for In solutes to occupy the Ga(2) site.

At elevated temperature, FeGa$_3$ spectra exhibited damping as shown in Figure 3. This is naturally attributed to relaxation due to diffusional motion of the probe atoms. Spectra could be reproduced well using Eq. 2. Best-fit values obtained for the relaxation parameter are shown as an Arrhenius plot in Fig. 4. As can be seen, the temperature dependence of the relaxation parameter is consistent with a single-step activation process. A linear fit of this temperature dependence yields an activation energy of 1.6±0.1 eV.

As described below, different Cd jump pathways among Ga(2) sites result in different types of EFG reorientations. In principle, the different types of reorientations will produce damping and shifts of PAC signal harmonics that are subtly different, and under favorable conditions, contributions from the two types will be resolvable if spectra are fitted with a stochastic model of fluctuating EFGs. This can provide more insight about how Cd moves beyond what is possible using the empirical analysis.

4  Stochastic Model Analysis

As shown below, a stochastic model of fluctuations among the Ga(2) sites requires three parameters for the EFG states: the EFG strength or main principal component $V_0$, the asymmetry parameter $\eta$, and one directional angle to describe EFG orientation. The model also requires two distinct rates of change from one EFG orientation to another. It turns out that due to interdependencies of model parameters, it was possible to determine only one of the reorientation rates with meaningful certainty.

## 4.1 Model details

FeGa$_3$ has eight crystallographically equivalent Ga(2) sites. A PAC probe at any of the sites will experience the same EFG strength and asymmetry parameter; however, orientations of EFG axes vary among the sites. Each Ga(2) site has a mirror plane perpendicular to a vector in a (1,±1,0) direction of the crystal, where the directions ($x,y,z$) are defined by the conventional lattice vectors (with the $z$ coordinate along the $c$ axis of the crystal). This constrains one of the principal EFG axes to be perpendicular to the mirror plane; the other two EFG axes must be perpendicular to each other and lie within the mirror plane. Mirror planes in the space group require that EFG axes within a mirror plane have only two unique orientations. As a result, there are only four possible EFG orientations at Ga(2) sites, labeled A through D in Fig. 1. It should be noted that A and C sites are interchanged and B and D sites are interchanged in adjacent Ga(2) planes along the $c$-axis.

EFG reorientations will occur when probes jump among the A, B, C, and D sites. Depending on the underlying diffusion mechanism, there are four different single-step jumps among Ga(2) sites. There are two types of jumps within an $a$-$b$ plane: along a rhombus (or square) edge shown in Fig. 1 with rate $w_1$ and across the short diagonal of a rhombus with rate $w_2$. There are two out-of-plane jumps to "interchanged" sites with slightly different jump lengths; the different jump distances likely will result in two different jump rates $w_3$ and $w_3'$. The single-step jumps for an atom starting at an A-site are shown in Fig. 5.

Without consideration of symmetry, there are 12 EFG reorientation rates, $r_{i \to j}$ where $i$ denotes the source site's EFG and $j$ denotes the destination EFG. Inspection of Fig. 5 shows $r_{A \to B} = 2w_1$, $r_{A \to C} = w_2 + w_3 + w_3'$, and $r_{A \to D} = 2w_1$. Jumps A→B and A→D result in the two possible changes in EFG that occur when the component perpendicular to the mirror plane reorients by 90°. As can be seen, the rates of change to either configuration are equal, and it is convenient to define the symbol $r_1$ to represent the total rate at which the component perpendicular to a mirror plane reorients; that is, $r_1 = r_{A \to B} + r_{A \to D}$. The other jump pathways do not lead to changes in the perpendicular-component of the EFG. It is convenient to define a second rate $r_2$ as the total rate at which the non-perpendicular component reorients; that is, $r_2 = r_{A \to C}$. Similar consideration of jumps originating at sites with other EFG orientations allows one to express the various $r_{i \to j}$ in terms of just $r_1$ and $r_2$, as expected from symmetry.

## 4.2 Theoretical prediction of EFG orientation

Exploratory fits revealed a strong correlation between EFG fluctuation rates and orientations of EFGs in the crystal system. Without additional information about EFG orientations, it was not possible to obtain meaningful fluctuation rates. Therefore, the EFG experienced by Cd at a Ga(2) site was calculated using density functional theory (DFT) under the APW+lo basis set [34] as implemented in the WIEN2k code [35]. The Wu and Cohen generalized gradient approximation (GGA) [36] was used as the exchange-correlation functional, and calculations were carried out without spin polarization. A Ga(2) atom was replaced by Cd in a 2×2×2 supercell of FeGa$_3$, centered on the Cd. Calculations were run to self-consistency, using the tetrahedron method [37] to perform integrations in the reciprocal space sampled by 2 $k$-points in the reduced first Brillouin zone. Lattice parameters and atomic positions were allowed to relax until forces were less than 0.02 eV/Å with a basis set size determined using $R_{MT}K_{max} = 7.0$. After that, a final calculation of the EFG was performed with $R_{MT}K_{max} = 9.5$.

Using the conventional notation with $V_{zz}$, $V_{yy}$, and $V_{xx}$ denoting the largest to smallest magnitude components of the EFG in terms of its principal axes, the $V_{xx}$ component is the one perpendicular to the Ga(2)-site mirror plane while the $V_{yy}$ and $V_{zz}$ components lie in the mirror plane. The $V_{yy}$ axis forms a 6.5°-angle with respect to the $c$-direction, tilted in the direction of the closest Fe atom in the same mirror plane in which the Cd sits. The $V_{zz}$ axis, then, is at 83.5° with respect to the $c$-direction. $V_{zz}$ was found to be $-21.5 \times 10^{21}$ V/m$^2$; using $Q=0.83$ b for Cd [38], this gives $\omega_Q = 66$ Mrad/s, which differs from the experimental value at room temperature by about 30%. The asymmetry parameter was found to be 0.7873 as opposed to the experimental value of 0.939. Given these discrepancies between calculated and observed values for $V_{zz}$ and $\eta$, it is necessary that they be treated as adjustable parameters within the fluctuation model. The most important contribution of the DFT calculation is to establish that $V_{xx}$ is perpendicular to the Ga(2)-site mirror plane. This means that only one additional adjustable parameter for the EFG is needed: $\theta$, which was chosen to be the angle that $V_{zz}$ makes with the $c$-axis.

### 4.3 Fit results

Adjustable parameters in the stochastic model are the strength of the EFG, $V_0 \equiv |V_{zz}|$, asymmetry parameter $\eta$; the angular deviation of the $V_{zz}$ axis from the $c$-direction, $\theta$; the total rate at which the $V_{xx}$ orientation changes, $r_1$; and the rate at which $V_{yy}$ and $V_{zz}$ axes interchange without a change in $V_{xx}$, $r_2$. The information based on these parameters needed to construct the Blume matrix for evaluation of Eq. 3 is given in the Appendix.

Even with the restriction that the $V_{xx}$ principal axis lies perpendicular to the Ga(2) mirror plane, fits in which all five model parameters were allowed to adjust resulted large uncertainties in reorientation rates. Reasonable uncertainties could be obtained by holding $\theta$ at a fixed value in fits; however, the value 1.457 rad predicted by DFT should be treated with caution given the discrepancy between DFT and experimental values for $\omega_Q$ and $\eta$. The sensitivity of $r_1$ and $r_2$ to choice of $\theta$ was examined over a range of values around 1.457 rad. Fit results obtained for the 550 °C spectrum are shown in Table 1. As can be seen, $r_2$ is highly dependent on the choice of $\theta$ whereas $r_1$ is relatively insensitive to changes in $\theta$ over the range of interest.

Given that $\theta$ is uncertain both because of the known discrepancies between DFT and experiment in values for $\omega_Q$ and $\eta$ and because $\theta$ is likely to change with temperature due to thermal expansion and increasing atomic vibration amplitudes [39], it does not appear possible to obtain meaningful values for $r_2$ without additional experiments on single crystals to determine EFG orientation as a function of temperature. In the meantime, it is possible to obtain useful values for $r_1$. Best-fit values for $r_1$ obtained in fits with $\theta$ held fixed at the predicted value 1.457 rad are shown in Fig. 4. The temperature dependence of $r_1$ is consistent with a single step activation process, and a linear fit of this temperature dependence yields an activation energy of 1.8±0.1 eV.

## 5   Discussion

The strong preference of In for the Ga(2) site means that it is not possible to obtain a difference in site energies between inequivalent sublattices of the same element like it was in the $Pd_3Ga_7$ and $Al_3Ti$ systems. In this way, In in the $FeGa_3$ system more closely resembles the distribution of In in $\beta$-Mn, which has two inequivalent Mn-sites of which In only appears on one throughout the temperature range studied [14].

A heuristic explanation for the strong site preference experienced by In in $FeGa_3$ is that occupation of the Ga(2) site minimizes contact between In and Ga and maximizes contact with Fe-atoms. In other work, the solubility of In in solid Ga was found to be extremely low: of the order $10^{-12}$ [32], suggesting that it is energetically favorable for In to avoid Ga as much as possible in a solid. The Ga(2) site in $FeGa_3$ has more near neighbor Fe atoms than the Ga(1) site, and thus less contact with neighboring Ga.

The choice to consider only single-step jump pathways among Ga(2) sites as shown in Fig. 5 is based on the assumption that diffusion occurs by a vacancy mechanism in which Cd remains on the Ga(2) sublattice. The EFG reorientation model can be extended easily to include other diffusion mechanisms for which Cd passes through Ga(1) sites or even Fe sites in the event that the Fe vacancy concentration is much larger than the Ga vacancy concentration [40].

Unfortunately, because it was not possible to obtain spectra for $FeGa_3$ samples rich in Ga, it was not possible to study the composition dependence of the relaxation parameter $\lambda$ or jump rate $r_1$ and by extension learn partial information about the operative diffusion mechanism as it was in the series of rare earth tri-indides [9, 10]. With activation energies of 1.6 and 1.8 eV respectively for $\lambda$ and $r_1$, the jump rate of Cd in $FeGa_3$ is on the high end of the range of activation energy (0.5-1.6 eV) for Cd jumps in other systems [41,12]. The large activation energy is suggestive that cohesion, or binding, is stronger in the $FeGa_3$, which would be consistent with the view that $FeGa_3$ has a larger-than-usual covalence in the interactions between atoms.

The present work is of special interest in the context of studying stochastic fluctuations of EFGs. Usually, fluctuations occur due to a change in EFG strength or a change in $V_{zz}$ orientation. As an example, changes in only EFG strength were observed for Cd-donor complexes in Si [42]. The most common fluctuation involving reorientation of $V_{zz}$ is characterized by the XYZ model [20,21,43] and has been observed experimentally in $L1_2$-structured compounds [9,10, 12,41]. The present case is unusual in that the reorientation of the $V_{xx}$ axis (upon which $r_1$ is sensitive) is of greater significance than the reorientation of the $V_{zz}$ axis (upon which $r_2$ depends).

## 6   Summary

Results were reported of $^{111}$In PAC experiments carried out primarily in $FeGa_3$ but also in $RuIn_3$. A single PAC signal was observed for $FeGa_3$ with hyperfine parameters $\omega_1 = 513.8\pm0.1$ Mrad/s and $\eta = 0.939\pm 0.002$ at room temperature. It was identified as substitutional In on a Ga(2) site. Two signals were observed for $RuIn_3$ with parameters $\omega_1 = 415.8\pm 0.4$ M/rad with $\eta = 0.015\pm0.015$ and $\omega_1 = 533.9\pm0.8$ Mrad/s with $\eta = 0.9316\pm0.0006$ at room temperature corresponding to In(1) and In(2) sites, respectively. As for solute site occupation behavior, In strongly favors the Ga(2) site. No signal that could be attributed to In on a Ga(1) site was observed, and it is therefore not possible to determine the difference in site occupation energies.

At elevated measurement temperature, PAC spectra exhibited damping due to nuclear relaxation attributed to jumps of Cd among Ga(2) sites. Using a stochastic model for fluctuating EFGs, it was possible to partially resolve rates of two distinct EFG reorientation processes. The total rate of jump processes that do not lead to reorientation of the $V_{xx}$ EFG axis could not be determined with reasonable precision. The total rate of processes that lead to a rotation of $V_{xx}$ by 90° could be, and it was found to be thermally activated with an activation enthalpy of 1.8±0.1 eV.

**Acknowledgement**

This work was supported in part by NSF grants DMR 06-06006, 09-04096, and 14-10159.

**Appendix**

In order to construct the Blume matrix, one must specify the EFGs and rates of transition among the EFGs in the stochastic model. For jumps among the Ga(2) sites in FeGa$_3$, the four EFGs are given by

$$V_A = \begin{pmatrix} \frac{V_{xx} + C^2 V_{yy} + S^2 V_{zz}}{2} & \frac{V_{xx} - C^2 V_{yy} - S^2 V_{zz}}{2} & (-V_{yy} + V_{zz})\frac{CS}{\sqrt{2}} \\ \frac{V_{xx} - C^2 V_{yy} - S^2 V_{zz}}{2} & \frac{V_{xx} + C^2 V_{yy} + S^2 V_{zz}}{2} & (V_{yy} - V_{zz})\frac{CS}{\sqrt{2}} \\ (-V_{yy} + V_{zz})\frac{CS}{\sqrt{2}} & (V_{yy} - V_{zz})\frac{CS}{\sqrt{2}} & S^2 V_{yy} + C^2 V_{zz} \end{pmatrix}$$

$$V_B = \begin{pmatrix} \frac{V_{xx} + C^2 V_{yy} + S^2 V_{zz}}{2} & \frac{-V_{xx} + C^2 V_{yy} + S^2 V_{zz}}{2} & (-V_{yy} + V_{zz})\frac{CS}{\sqrt{2}} \\ \frac{-V_{xx} + C^2 V_{yy} + S^2 V_{zz}}{2} & \frac{V_{xx} + C^2 V_{yy} + S^2 V_{zz}}{2} & (-V_{yy} + V_{zz})\frac{CS}{\sqrt{2}} \\ (-V_{yy} + V_{zz})\frac{CS}{\sqrt{2}} & (-V_{yy} + V_{zz})\frac{CS}{\sqrt{2}} & S^2 V_{yy} + C^2 V_{zz} \end{pmatrix}$$

$$V_C = \begin{pmatrix} \dfrac{V_{xx}+C^2V_{yy}+S^2V_{zz}}{2} & \dfrac{V_{xx}-C^2V_{yy}-S^2V_{zz}}{2} & (V_{yy}-V_{zz})\dfrac{CS}{\sqrt{2}} \\ \dfrac{V_{xx}-C^2V_{yy}-S^2V_{zz}}{2} & \dfrac{V_{xx}+C^2V_{yy}+S^2V_{zz}}{2} & (-V_{yy}+V_{zz})\dfrac{CS}{\sqrt{2}} \\ (V_{yy}-V_{zz})\dfrac{CS}{\sqrt{2}} & (-V_{yy}+V_{zz})\dfrac{CS}{\sqrt{2}} & S^2V_{yy}+C^2V_{zz} \end{pmatrix}$$

$$V_D = \begin{pmatrix} \dfrac{V_{xx}+C^2V_{yy}+S^2V_{zz}}{2} & \dfrac{-V_{xx}+C^2V_{yy}+S^2V_{zz}}{2} & (V_{yy}-V_{zz})\dfrac{CS}{\sqrt{2}} \\ \dfrac{-V_{xx}+C^2V_{yy}+S^2V_{zz}}{2} & \dfrac{V_{xx}+C^2V_{yy}+S^2V_{zz}}{2} & (V_{yy}-V_{zz})\dfrac{CS}{\sqrt{2}} \\ (V_{yy}-V_{zz})\dfrac{CS}{\sqrt{2}} & (V_{yy}-V_{zz})\dfrac{CS}{\sqrt{2}} & S^2V_{yy}+C^2V_{zz} \end{pmatrix}$$

where $C \equiv \cos\theta$, $S \equiv \sin\theta$, and $V_{xx}$ is taken to be perpendicular to the Ga(2)-mirror plane as determined by DFT calculations. The transition rate matrix is given by

$$R = \begin{pmatrix} -2r_1+r_2 & r_1 & r_2 & r_1 \\ r_1 & -2r_1+r_2 & r_1 & r_2 \\ r_2 & r_1 & -2r_1+r_2 & r_1 \\ r_1 & r_2 & r_1 & -2r_1+r_2 \end{pmatrix}$$

where $r_1$ and $r_2$ are the reorientation rates defined in section 4.1.

Figures

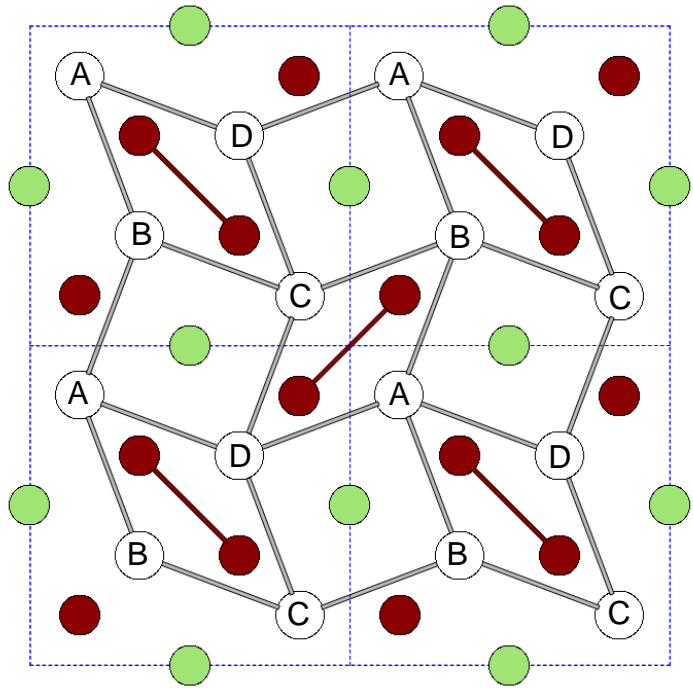

Figure 1. Crystal structure of FeGa$_3$. Four unit cells in the a-b plane are shown. Lightly shaded atoms (green online) are Ga(1) and positioned at fractional positions 0, 1/2, and 1 along the *c* axis (into the page). Darkly shaded atoms (red online) are Fe: those connected by bars oriented from the upper left to the lower right are located at fractional positions 0 and 1 along the c axis; the other Fe atoms are located at c/2. The unshaded atoms are Ga(2) and are located near planes at the c/4 and 3c/4 positions (with the c/4 positions visible). The labels on the Ga(2) atoms indicate orientations of electric field gradients at those sites, as discussed in section 4.

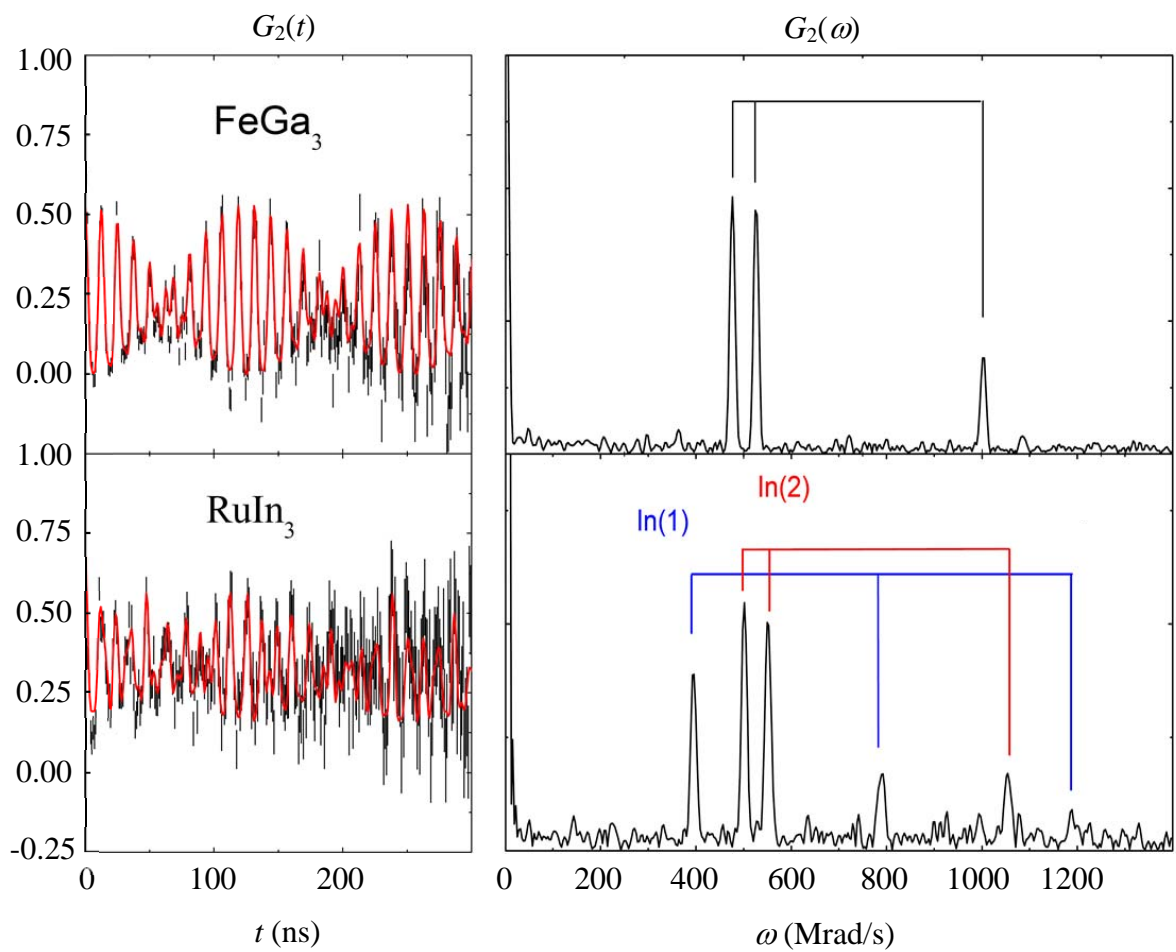

Figure 2. $G_2(t)$ and Fourier transform for FeGa$_3$ taken at 400 °C (top) and for RuIn$_3$ taken at 500 °C (bottom).

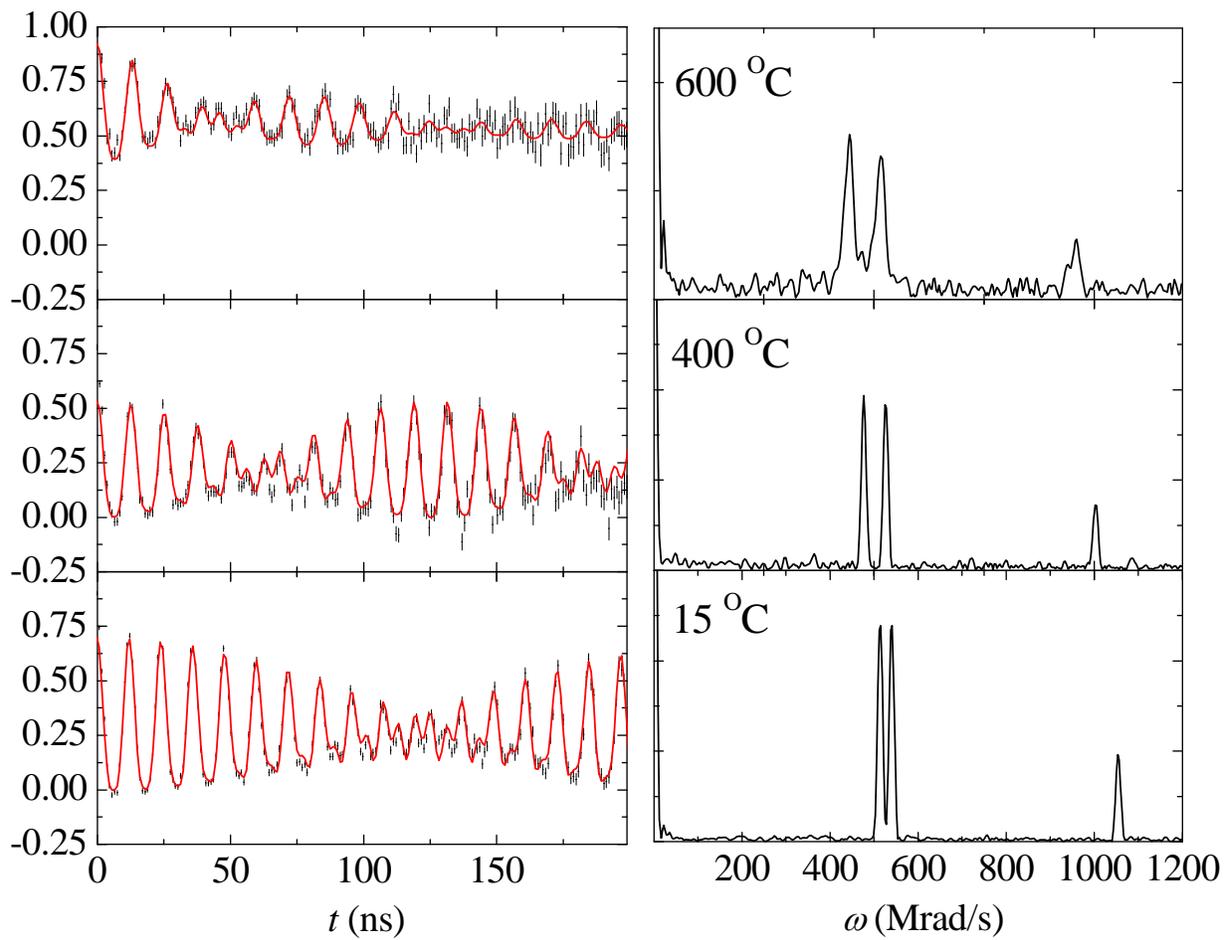

Figure 3. $G_2(t)$ and Fourier transforms for FeGa$_3$ taken at indicated temperatures.

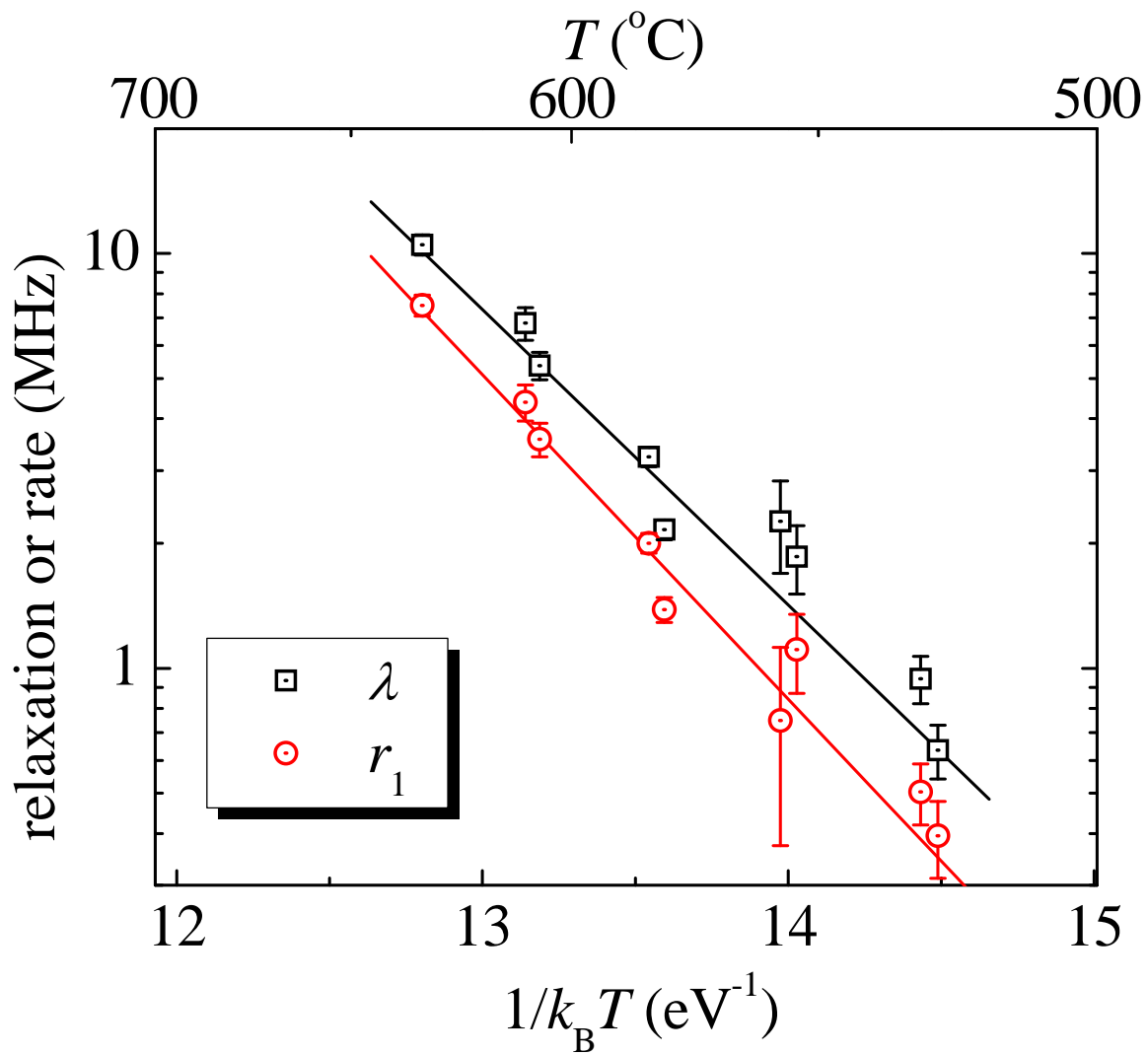

Figure 4. Arrhenius plot of relaxation rate $\lambda$ obtained by fits using Eq. 2 (squares) and of rate $r_1$ obtained by fits using Eq. 3 (circles) with best fit lines shown.

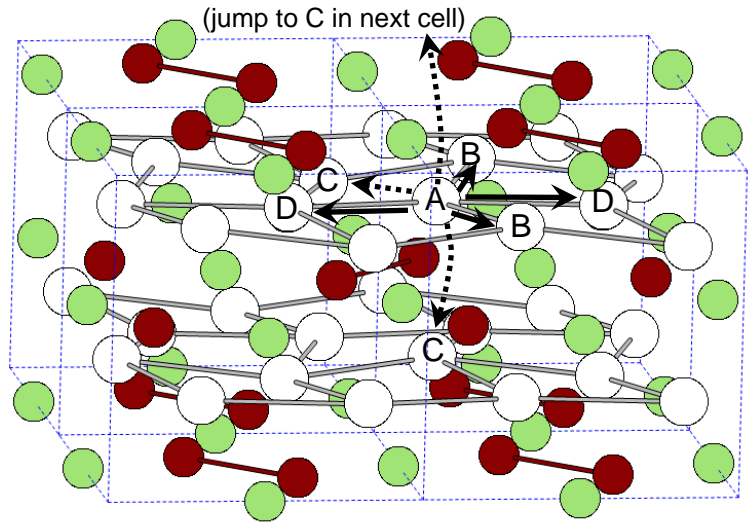

Figure 5. Crystal structure of FeGa$_3$ with possible jump vectors from a Ga(2) site with EFG type A. There are two types of jumps within an a-b plane: along a rhombus edge with rate $w_1$ (solid arrows) and across a rhombus with rate $w_2$ (straight dashed arrow). There are two out-of-plane jumps to C-sites with slightly different jump lengths (curved dashed arrows) with jump rates $w_3$ and $w_3'$.

Table

Table 1. EFG reorientation rates $r_1$ and $r_2$ obtained from fits to the spectrum collected at 550 °C for different fixed values of $\theta$.

| $\theta$ (rad) | $r_1$ (MHz) | $r_2$ (MHz) |
|---|---|---|
| 1.30 | 1.6±0.4 | 0.4±0.7 |
| 1.35 | 1.5±0.4 | 0.8±0.9 |
| 1.40 | 1.5±0.4 | 1.5±1.4 |
| 1.45 | 1.5±0.4 | 2.8±2.6 |
| 1.50 | 1.7±0.4 | 7.4±7.8 |